\documentstyle [prl,aps]{revtex}
\newcommand{\la}[1]{\label{#1}}

\newcommand{\Ho}{{\cal H}}
\newcommand{\Tw}{{\cal T}}
\newcommand{\Wr}{{\cal W}}

\newcommand{\n}{{\bf n}}

\newcommand{\be}{\begin{equation}}
\newcommand{\ee}{\end{equation}}
\newcommand{\ba}{\begin{eqnarray}}
\newcommand{\ea}{\end{eqnarray}}
\newcommand{\bastar}{\begin{eqnarray*}}
\newcommand{\eastar}{\end{eqnarray*}}

\newskip\Humongous \Humongous=0pt plus 1000pt minus 1000pt

\newif\ifdtup

\relax
\begin{document}
\title  {ASPECTS OF DUALITY AND CONFINING STRINGS \\ }
\bigskip

\author{Mauri Miettinen$^{*}$, 
Antti J. Niemi$^{* \dagger}$ and Yuri Stroganov$^{\#}$} \vskip 0.3cm
\address{$^*$Department of Theoretical Physics, Uppsala University,
Box 803, S-75108 Uppsala, Sweden \\
$^{*}$The Mittag-Leffler Institute, Aurav\"agen 
17, S-182 62 Djursholm, Sweden \\ 
$^{\dagger}$Helsinki Institute of Physics, P.O. Box 9, FIN-00014 
University of Helsinki, Finland \\
$^{\#}$ Institute for High Energy Physics, 
Protvino, Moscow Region, Russia \\ 
$^{\#}$ Research Institute for Mathematical Sciences, 
Kyoto University, Kyoto 606, Japan \\ \vskip 0.2cm 
{\scriptsize \bf niemi@teorfys.uu.se\hskip 1.0cm 
strogano@kurims.kyoto-u.ac.jp } 
\\ \vskip 0.2cm
}

\maketitle

\begin{abstract}
We inspect the excitation energy spectrum of a confining 
string in terms of solitons in an effective field theory 
model. The spectrum can be characterized by a spectral 
function, and twisting and bending of the string
is manifested by the invariance of this function under
a duality transformation. Both general considerations
and numerical simulations reveal that asymptotically
the spectral function can be approximated by a 
simple rational form, which we propose becomes exact 
in the Yang-Mills theory.
\end{abstract}

\widetext
\bigskip

The ground state energy of a long confining string
is proportional to its length $L$. When $L$
decreases there will be corrections, and in QCD   
we expect for the ground state energy of the confining
flux \cite{kuti}
\begin{equation}
E_0(L) \ = \ \sigma \cdot L \ + \ \varepsilon \ + \ 
\frac{c}{L} \ + \ {\cal O}(\frac{1}{L^2})
\la{lusher}
\end{equation}
Here $\sigma$ is the string tension,
$\varepsilon$ the intercept, and $c = \frac{\pi}{12}$ 
is a universal constant, the Casimir energy of zero-point 
fluctuations that can be computed from 
the Nambu-Goto action in the Gaussian 
approximation \cite{kuti}. Unfortunately, 
a simple Nambu-Goto action becomes 
insufficient when we attempt to
describe the excitation energy spectrum.
Now there are additional contributions 
that can lead to large deformations even when 
the underlying displacements remain tiny. 
For example, a small torsional rotation 
around the axis of a long string can cause 
distant cross-sections to rotate through 
large angles. Similarly, if a long string is 
slightly bent its ends can move over considerable 
distances.  

In the present Letter we propose an effective 
field theory approach to describe the 
excitation energy spectrum of a confining string.
The string appears as a soliton in the effective 
field theory, which enables us to account 
for contributions such as bending and twisting 
at the (semi)classical level. This provides
us with an appropriate background, to systematically
investigate quantum corrections.
Quantitatively, our arguments are based on  
the classical action \cite{fadde}
\be
\int d^4x \, \left[ \,
(\partial_\mu {\n})^2 \, + \, \frac{1}{4e^2} 
( \n \cdot  \partial_\mu \n \times 
\partial_\nu \n )^2 \, \right]
\la{lagr}
\ee
The order parameter $\n$ is a three-component 
unit vector $\n \cdot \n = 1$ and $e^2$ is a
coupling constant that determines the length scale.
This action is known to support closed knotted strings as 
solitons \cite{nature} - \cite{bs}.
According to \cite{prl}, it also
represents a universality class
that describes SU(2) Yang-Mills theory 
in the long wavelength
limit. Indeed, (\ref{lagr}) does emerge from the 
Yang-Mills theory in a derivative expansion \cite{edwin}. 
Thus it becomes natural to use its string solutions to 
inspect properties of the confining flux
in the Yang-Mills theory. 
However, ultimately we wish to employ universality to 
propose that our results remain valid even beyond 
the steps that lead to (\ref{lagr}). For this we  
shall combine duality arguments with a familiar problem in 
the classical theory of elasticity, the evaluation 
of the free energy that describes elastic deformations 
of a thin rod \cite{landau}. At least 
to the extent our results parallel 
those in the classical theory 
of elasticity, we expect their validity to reach 
beyond the specific details of (\ref{lagr}).

Classically there are exactly two intrinsically 
small deformations of an elastic rod that can be accompanied by 
large global displacements, twisting about
its axis and bending \cite{landau}.
When a straight elastic  rod is twisted around 
its axis, each transverse section becomes rotated by 
some relative angle.  
We align the coordinates so that the axis of the rod
coincides with the $z$-axis and denote by $\tau$ 
the (local) twist angle, defined as the angle of 
rotation per unit length of the rod around its axis. 
For a constant $\tau$
a rod with length $L$ acquires a total 
twist $\psi = \tau L$ along its length. Combining 
(\ref{lusher}) with the results in \cite{landau} we then
expect that the free energy of a straight, twisted 
confining string has the form 
\be
E_{0} \, + \, E_{twist} \ = \ \sigma \cdot L \, 
+ \, \varepsilon \, + \, 
(2 \pi^2 {\cal C} + \frac{c}{\Tw^2}) \cdot \frac{\Tw^2}{L} 
\, + \, {\cal O}(\frac{1}{L^2})
\ = \ \sigma |\Tw| \cdot \left( \, \lambda \, + \, \epsilon \, + \,
\frac{a}{\lambda} \, \right) \ + \
{\cal O}(\frac{1}{L^2})
\la{flux1}
\ee 
Here $\Tw = \frac{1}{2\pi} \psi$ counts the  
number of twists around the axis, ${\cal C}$ 
is a form-factor that characterizes the string, and
the variable $\lambda = L/|\Tw|$ denotes length per 
twist. We note that (\ref{flux1})
exhibits a $\lambda \to a/\lambda$ duality, reminiscent 
of the T-duality in (super)string theories. 
When the number of twists $\Tw$ remains fixed
the self-dual point $\lambda = \sqrt{a}$ minimizes 
the ensuing contribution which yields  
\be
E_{min}(Q) \ \approx \ 
2\sqrt{a} \cdot \sigma |\Tw| \ + \ \varepsilon
\ + \ {\cal O}(\frac{1}{L^2})
\la{flux2}
\ee 

In the field theory model (\ref{lagr}) a  
straight twisted string is described 
by an axially symmetric 
configuration, in cylindrical coordinates
\be
\n(r,\varphi,z) \ = \ \left( \matrix{ \sin 
( k \varphi + \tau z )
\, \sin \theta(r) \cr
\cos ( k \varphi + \tau z ) \, \sin \theta (r) \cr
\cos \theta(r) } \right)
\la{an}
\ee
where $k$ is an integer that counts the degeneracy. 
We substitute (\ref{an}) into (\ref{lagr}), and
by defining $\rho = e r$ a dimensionless variable,   
we find for the energy of a string 
with length $L$
\be
\frac{E}{2\pi} \ = \ 
\left\{ \, \int\limits_0^{\infty} 
\rho d \rho\, \left( \theta_{\rho}^2 + k^2 \cdot \left[ 
\frac{1 + \theta_\rho^2}{\rho^2}
\cdot \sin^2 \theta \right] \right) \, \right\} \cdot L 
\ + \ \frac{4 \pi^2 \Ho^2}{k} 
\cdot \left\{ \, \int\limits_0^{\infty} 
\rho d \rho \, ( 1 + \theta_\rho^2) \cdot \sin^2 \theta ) \, 
\right\} \cdot \frac{e^2}{L}
\la{twistE}
\ee
We draw attention to the similarity in the 
functional forms of (\ref{flux1}) 
and (\ref{twistE}) which can  
be used to relate the 
parameters $\sigma$ and ${\cal
C}$ to integrals over $\theta(\rho)$. 
Here we have introduced the Hopf invariant
$\Ho$, explicitely with $F = (\n, d\n \wedge d\n) = dA$ 
\[
\Ho \ = \ \frac{1}{2\pi} \int d^3x \, F \wedge A \ = \ 
\frac{1}{2\pi} k \tau L \ = \ \frac{k}{2\pi} \psi
\]
so that for a straight string the Hopf invariant
reduces to the number of twists $\Ho=\Tw$
that a configuration with 
length $L$ makes around the $z$-axis (including the 
multiplicity $k$).

The profile $\theta(\rho)$ that minimizes the energy 
(\ref{twistE}) solves the pertinent
Euler-Lagrange equation. Since this equation depends
nontrivially on the various parameters in (\ref{twistE}), 
we expect {\it a priori} that the solution also
attains a nontrivial, nonlinear dependence on the 
parameters. In particular we expect that for an actual 
minimum energy configuration the functional form of
(\ref{twistE}) in $L$ does not stand. 
 
We have performed an exhaustive numerical investigation of 
the parameter dependence for the configuration  
$\theta(\rho)$ that minimizes the energy (\ref{twistE}). 
The result is described by 
a spectral function $f(k^2,\lambda)$. It depends on
the integer $k^2$ and the dimensionless combination 
$\lambda = (Le)/(2\pi |\Ho|)$ 
which measures length per Hopf invariant.
Quite surprisingly, we find that the functional 
form (\ref{flux1}) {\it 
persists}: We find that to a 
very high degree of accuracy the energy of an actual 
straight string solution is interpolated by 
a simple rational spectral function
\be
{ E_{min}(k^2,\lambda) \over 4 \pi^2 e } \, = \, |\Ho| \cdot
f(k^2 , \lambda) 
\, \approx  \, 
|\Ho| \cdot \left( \, a \lambda \, 
+ \, b \, + \, \frac{c}{\lambda} \, \right)
\la{interp}
\ee
where $a,b,c$ are $k^2$ dependent numerical 
coefficients.
In figure 1 we plot (\ref{interp}) for $k=1$, 
together with several numerically computed 
values of the energy. The high degree of accuracy 
of the rational spectral function (\ref{interp}) 
over a wide range of values in $\lambda$ and $k^2$ 
leads us to propose that this simple rational
form might actually be exact for 
the straight string solutions
of (\ref{lagr}) (or at least to a slight perturbation
of (\ref{lagr}) within its universality class).

We note that (\ref{interp}) has 
a definite, manifest modular 
structure as it should. Namely, suppose 
we split a string into two so that the total
energy remains intact. 
Obviously, the spectral form 
of the energy for both of the resulting 
strings should also coincide with (\ref{interp}). 
Since the Hopf invariant $|\Ho|$ is additive and $\lambda$ 
denotes length per Hopf invariant, we conclude
that (\ref{interp}) indeed does have the 
requisite property under splitting and joining.  

The definite rational form (\ref{interp}) 
leads to additional observations. For this we 
first note that if $\lambda =
\lambda_c$ minimizes (\ref{interp}), in parallel with
(\ref{flux1}) the spectral function in (\ref{interp})
exhibits a $\lambda \to 
\lambda^2_c /\lambda$ duality. In order to interpret this 
duality we consider a string with Dirichlet-type 
boundary conditions that clamp the ends which keeps 
$\Ho$ intact. According to (\ref{interp}) 
the string has a tendency to adjust its
length $\lambda$ towards $\lambda = \lambda_c$
{\it i.e.} towards the self-dual point of the 
spectral function that minimizes the energy:
When $\lambda < \lambda_c$ the value of $\lambda$ has a 
tendency to increase, but when 
$\lambda > \lambda_c$ there is a tendency for $\lambda$
to decrease. 

We recall the relation between Hopf 
invariant $\Ho$, twist $\Tw$ and 
writhe $\Wr$ \cite{hopf}
\be
\Ho \ = \ \Tw \ + \ \Wr  
\la{h}
\ee
From this we conclude that for $\lambda \not= \lambda_c$
the string has a tendency to {\it supercoil}:
When $\lambda < \lambda_c$ it becomes energetically
favourable for the string to dilate. This
tends to decrease twist $\Tw$, hence there is a
tendency to increase writhe $\Wr$. Similarly, when $\lambda > 
\lambda_c$ it becomes favourable to contract which can 
lead to an increase in twist at the expense of writhe. 
As a consequence the $\lambda \to \lambda^2_c /\lambda$ 
duality of the rational spectral function becomes related 
to supercoiling. It can map a supercoiled configuration 
with an excess amount of twist to a (dual) supercoiled 
configuration with an excess amount of writhe but with 
an equal amount of energy: Since the Hopf invariant $\Ho$ 
does not change when we exchange $\Tw \leftrightarrow \Wr$,
we have a twist-writhe duality 
mapping that exchanges $\Tw \leftrightarrow \Wr$ and 
sends $\lambda \to \lambda_c^2 /\lambda$ but leaves the
the spectral function intact. The energy (\ref{interp})
is invariant under this twist-writhe duality 
transformation, it determines a  
symmetry in our effective field theory \cite{note}. 

The previous arguments are based on the inspection
of a straight string. But a configuration with a 
nontrivial writhe can not be straight, it must be bent. 
For this we recall from classical theory of elasticity that 
the free energy for bending of a thin elastic rod,
uniformly bent over a length $L$, has the form $E
= \kappa L /R^{-2}$ with $R$ the (mean) 
radius of curvature and $\kappa$ a form factor 
that characterizes the rod \cite{landau}. 
Suppose we consider a straight confining string with Hopf 
invariant $\Ho$ and a constant rate of twist along the 
string. We bend the string slightly in a uniform and non-planar
manner, with a constant radius of curvature $R$. 
This leads to a decrease in twist and, since $\Ho$ remains
intact, to a nontrivial writhe which we describe by 
a constant rate of increment $\omega$ along the string. For
small $\omega$ the 
length $L(\omega)$ of the string scales in 
proportion to $\omega$ according to $L(\omega) 
= L_0 + c \cdot \omega R$, where $L_0$ is the (planar) 
component of the length that 
does not contribute to the writhe and 
$c$ is a numerical constant; for simplicity
we set $c=1$. When we combine the free energy for 
bending with the ensuing free energy (\ref{flux1}) 
for stretching and twisting, we conclude that to a leading order
the free energy of our string acquires the following
$\omega$ dependence,
\be
E(\omega) \, = \, \sigma L(\omega) \, + \ 
\alpha^2 \cdot \frac{(\Ho - \omega)^2}{L(\omega)} \, + \,
\beta^2 \cdot \frac{L(\omega)}{R^2}
\la{bendtwist}
\ee
with $\alpha$ and $\beta$ some parameters.

We first consider the minimum value of the
free energy (\ref{bendtwist}) when we 
minimize it {\it w.r.t.} the rate of 
increment $\omega$ by
keeping the Hopf invariant $\Ho$ fixed. We get  
\be
E_{min} \, = \, 2 (\rho - \alpha)\cdot \frac{\alpha}{R} 
\cdot \left( \, |\Ho| \, + \, \frac{L_0}{R} \, \right)
\la{minbent}
\ee
where we have defined $\rho^2 = \alpha^2 + \beta^2 + \sigma R^2$.
We observe that (\ref{minbent}) has the same functional form 
as (\ref{flux2}). This suggests in particular, that the 
underlying dual structure persists. For this we subject
the free energy to a minimization of $\omega$, with the 
condition that both the Hopf invariant $\Ho$ and the
total length of the string $L = L_0 + \omega R$ remain 
fixed. We find
\be
E_{min} \, = \, |\Ho| \cdot \frac{L}{S} \cdot
\left( \, \sigma \cdot \frac{S}{|\Ho|} \, + \, (\alpha
\beta)^2 
\cdot \frac{|\Ho|}{S} 
\, \right) 
\la{min2}
\ee
where $S^2 = \alpha^2 (L-L_0)^2 + \beta^2 L^2$.
The dual structure is now manifest. We also note
that both (\ref{minbent}) and (\ref{min2}) are 
consistent with the expected modular structure
under splitting and joining. 

In the field theory model (\ref{lagr}) a stable, bent  
and twisted confining string is (locally) described by a 
closed knotted soliton \cite{nature},
\cite{bs}. The radius of curvature of an actual soliton 
in general fails to be constant, but since the deviations 
from an average $R$ appear to remain small \cite{bs} 
we can expect the main features of (\ref{bendtwist})  
to persist. The energy of a closed knotted soliton should then 
admit an asymptotic expansion in terms of modular invariant
quantities such as Hopf invariant and length/average
value of the radius of curvature. In particular, we  
expect that the leading contribution in this expansion
has the manifestly twist-writhe dual functional 
form (\ref{interp}), (\ref{min2}) of the spectral representation, 
with the various parameters 
now integrals of $\n$ evaluated for the soliton. 
Since the soliton minimizes the energy it corresponds to 
the self-dual minimum energy point of the
spectral function. According to (\ref{interp}), (\ref{minbent}) 
this means that to the leading order in an asymptotic
expansion the energy spectrum of a closed knotted 
soliton admits the now-familiar functional form
\be 
E_{soliton} \ = \ \gamma \cdot \left( \, |\Ho| 
\, + \, \varepsilon \, \right) 
\la{torus3}
\ee 
with some parameters $\gamma$ and $\varepsilon$ \cite{com}. 
In particular, since $|\Ho| = 1,2,3, ... $ 
is an integer this energy spectrum is equipartitioned, 
$E_n =  \gamma \cdot (n + \varepsilon)$.
  
Due to the high complexity of the Euler-Lagrange equations 
of (\ref{lagr}) it becomes difficult to actually check the
accuracy of the asymptotic (\ref{torus3}), or to inspect the underlying 
dual structure. Numerical 
simulations are very demanding \cite{bs}, and at the moment 
we lack computational resources to perform 
exhaustive simulations. Thus we rely on 
the numerical results that have been recently published  
in \cite{bs}. There, the energy of various knotted 
solitons is evaluated for $Q_H = 1, ... , 8$. In 
figure 2 we compare the data in \cite{bs} 
to the predictions of (\ref{torus3}). 
When we account for the numerical uncertainties in the data
as described in \cite{bs}, we conclude that the agreement 
appears quite satisfactory. In particular, this 
suggest that for (\ref{lagr}) the dual 
twist-writhe structure is inherent.

\vskip 0.2cm
In conclusion, we have inspected properties of a 
confining string by describing it as a soliton in
an effective field theory. In particular, we have 
numerically studied the excitation energy spectrum 
of a straight string. We find that it can be described 
by a spectral function that admits a simple rational form, 
manifestly invariant under a duality transformation that 
relates twist and writhe. We have verified that the 
presently available 3D data \cite{bs} is also consistent 
with the simple rational form of the spectral function.
This suggests that (\ref{lagr}) is at least in the 
same universality class with a Lagrangian for which 
a manifestly dual, simple rational form of the spectral 
function is {\it exact}. It becomes natural 
to expect that our rational realization of the 
twist-writhe duality is distinct to an 
effective Lagrangian that accurately describes the 
properties of the confining flux  in the Yang-Mills theory.

\vskip 0.2cm
We thank L. Faddeev and E. Keski-Vakkuri
for numerous discussions,
and S. Nasir for comments. A.J.N. also thanks O. Viro,
and Yu. S. thanks G. Pronko for a useful discussion.
Our research has been partially supported by NFR 
Grant F-AA/FU 06821-308. The research by Yu. S. 
has also been supported by grants RBRF 98-01-00070 
and INTAS 96-690, and he thanks Uppsala University and
Helsinki Institute of Physics for hospitality.
This article reports collaboration, largely 
completed by March 2nd 1999 when Mauri Miettinen 
unexpectedly passed away just one month 
before his scheduled Thesis defence.


\vfill\eject
\begin{flushleft}
{\bf Figure Caption}
\end{flushleft}
\vskip 0.5cm

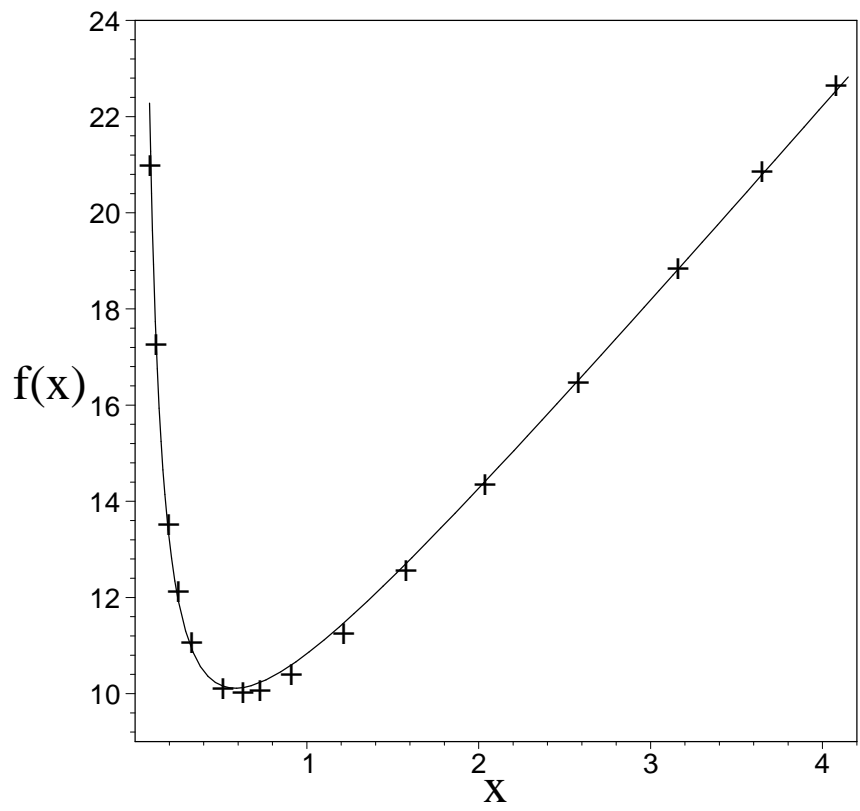
{\bf Figure 1:} A comparison of the rational function
(\ref{interp}) with the energy values of the line
solitons in (\ref{lagr}), (\ref{twistE}). 
The interpolation is in the
sense of least square, which yields $a \approx 4.15 ...$,
$b \approx 5.26 ...$ and $c \approx 1.42 ... $. The
agreement is consistent with finite lattice
size errors.

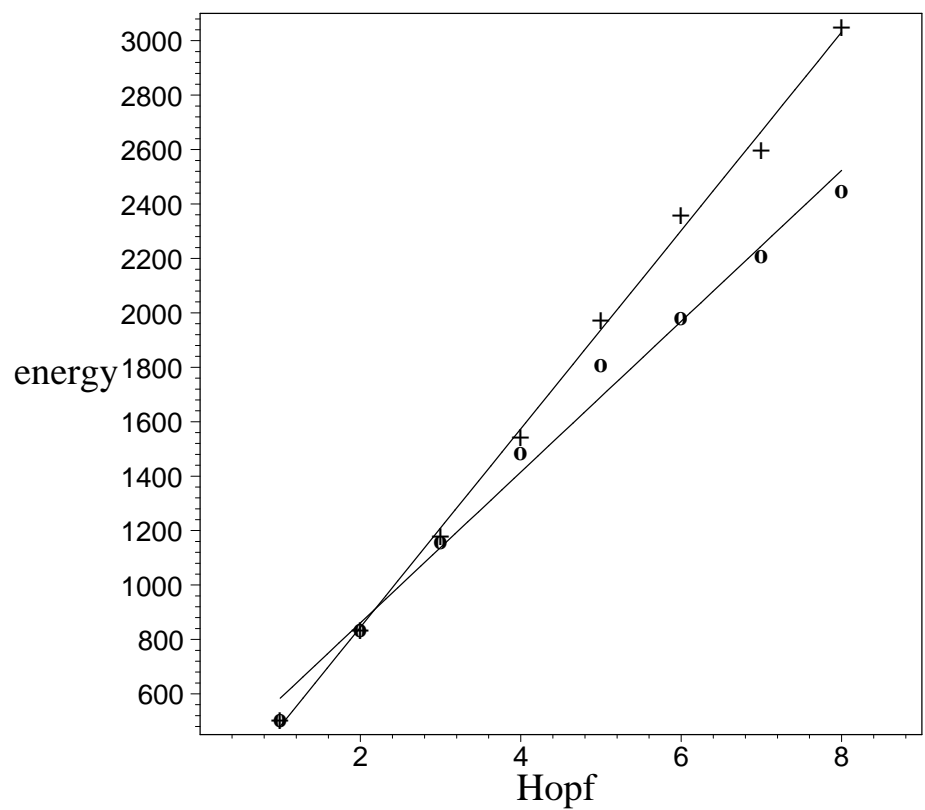
{\bf Figure 2:} Comparison of the linear trajectory
(\ref{torus3}) (see also \cite{com}) to numerics in \cite{bs}. 
Line/data with $+$ denotes the toroidal $\Ho = 
1, ...  ,8$ solutions with slope/intercept
364/116 and line/data with $\bf o$ the 
minimal energy $\Ho = 1,...,8$ solutions with slope/intercept
277/306. (We note that \cite{bs} considered
only interpolation with a vanishing intercept.)

\begin{thebibliography}{9}
\bibitem{note} We note that for large enough 
deviations $\lambda - \lambda_c $ the 
rational form (\ref{interp}) must eventually 
fail: A line string becomes unstable against splitting 
into closed configurations with 
nontrivial $\Ho$.
\end{thebibliography}
\end{document}